\documentclass[10pt]{IEEEtran}

\makeatletter
\def\ps@headings{%
\def\@oddhead{\mbox{}\scriptsize\rightmark \hfil \thepage}%
\def\@evenhead{\scriptsize\thepage \hfil \leftmark\mbox{}}%
\def\@oddfoot{}%
\def\@evenfoot{}}
\makeatother 
\usepackage{mathrsfs}
\usepackage{amsfonts}
\usepackage{array}   
\usepackage{amssymb}
\usepackage{amsmath}
\usepackage{graphicx}
\usepackage{multirow}  
\usepackage{amsthm}
\usepackage{graphicx}

\usepackage[lined,vlined,ruled,commentsnumbered]{algorithm2e}
\usepackage{epstopdf}
\usepackage{stmaryrd}
\usepackage{multirow,url}
\usepackage{color,cite}
\usepackage{slashbox}
\usepackage{bm}
\usepackage{indentfirst}
\usepackage{epsfig}
\usepackage{amsmath}
\usepackage[table]{xcolor}

\usepackage{tikz}
\usepackage{amssymb}

\newtheorem{theorem}{Theorem}
\newtheorem{lemma}{Lemma}

\newtheorem{definition}{Definition}
\newtheorem{proposition}{Proposition}

\theoremstyle{plain}

\ifodd 121
\newcommand{\rev}[1]{{#1}} 
 
\newcommand{\revhh}[1]{{#1}}

\newcommand{\comg}[1]{\textbf{\color{green} (COMMENT: #1)}}
\newcommand{\response}[1]{\textbf{\color{green} (RESPONSE: #1)}}
\else
\newcommand{\rev}[1]{#1}

\newcommand{\comg}[1]{}
\newcommand{\response}[1]{}
\fi

\newcommand{\revx}[1]{{\color{blue}#1}} 

\def\eq{\triangleq}

\def\N{\mathcal{N}}

\def\h{\textsc{h}}
\def\c{\textsc{c}}
\def\a{\textsc{a}}

\def\costc{\gamma_{\c}}
\def\costh{\gamma_{\h}}
\def\costhc{\gamma_{\h\c}}

\def\Cc{\phi_{\c}} 
\def\Ch{\phi_{\h}}

\begin{document}

\title{Incentive Design and Market Evolution of Mobile User-Provided Networks} 

\author{Mohammad Mahdi Khalili, Lin Gao, Jianwei Huang, and Babak Hossein Khalaj 
\vspace{-6mm}
\IEEEcompsocitemizethanks{
\IEEEcompsocthanksitem
Mohammad Mahdi Khalili and Babak Hossein Khalaj are with Department of Electrical Engineering, Sharif University of Technology, Iran,
E-mail: \{khalili\_ma@ee.sharif.edu, khalaj@sharif.ir\}; 
\IEEEcompsocthanksitem
Lin Gao and Jianwei Huang are with Department of
Information Engineering, The Chinese University of Hong Kong, HK,
E-mail: \{lgao, jwhuang\}@ie.cuhk.edu.hk; 
} }

\addtolength{\abovedisplayskip}{-1.5mm}
\addtolength{\belowdisplayskip}{-1.5mm}

\maketitle

\begin{abstract}
An operator-assisted user-provided network (UPN) has the potential  to achieve a low cost ubiquitous Internet connectivity,  without  significantly increasing the network infrastructure investment. 
In this paper, we consider such a network where the network operator encourages some of her subscribers to operate as mobile Wi-Fi hotspots (\emph{hosts}), providing Internet connectivity for other subscribers (\emph{clients}). 
We formulate the interaction between the operator and mobile users as a two-stage game. 
In Stage I, the operator determines the usage-based pricing and quota-based incentive mechanism for the data usage. 
In Stage II, the mobile users make their decisions about whether to be a host, or a client, or not a subscriber at all. 
We characterize how the users' membership choices will affect each other's payoffs in Stage II, and how the  operator optimizes her decision in Stage I to maximize her profit.  
\revx{Our theoretical and numerical results show that the operator's maximum profit increases with the user density under the proposed hybrid pricing mechanism, and the profit gain can be up to 50\% in a dense network comparing with a pricing-only approach with no incentives.}
\end{abstract}

\IEEEpeerreviewmaketitle

\section{Introduction}

%

%

\subsection{Background and Motivation}

The explosion of mobile data traffic and the proliferation of advanced mobile devices such as smartphones and tablets have led to a novel network connection paradigm, often known as \emph{User-provided Network (UPN)}, in
which  users act as micro-providers (hosts) through their personal devices  and provide Internet connections for others \cite{upn, UPN-magazine}. 
By exploiting the diversity of users' demands and crowdsourcing their Internet connectivities, the UPN can achieve a low-cost ubiquitous Internet connectivity  without the need of significant additional network infrastructure investment. 
Because of this, it has been recently introduced by 3GPP as a complementary communication and network scheme to existing infrastructure-based cellular network structures \cite{3GPP}. 


The commercial UPN services are becoming increasingly popular in recent years. 
Some of these services rely on \emph{fixed} hosts such as residential Femtocell and Wi-Fi access points (APs) \cite{FON, OpenSpark, Whisher, KeyWiFi, Telefonica}.\footnote{A typical commercial UPN example based on fixed hotspots is the Wi-Fi community network introduced by FON \cite{FON}, where each FON subscriber offers Internet access to other FON subscribers  through his residential Wi-Fi.} 
Other UPN models are more flexible, and rely on  \emph{mobile} host devices such as smartphones and customized portable devices \cite{Karma, OpenGarden}. 
One prominent commercial UPN example based on mobile hosts is the service launched by a recent Mobile Virtual Network Operator (MVNO)   Karma \cite{Karma}, which enables some of her subscribers to operate as Mobile Wi-Fi (MiFi) {hosts} (or simply called \emph{hosts}) and offer Internet access to other subscribers (called \emph{clients}). 
Another mobile UPN example is enabled by the crowdsourced networking software  OpenGarden \cite{OpenGarden}, where nearby mobile devices create a mesh network, and aggregate their Internet connections. 
\rev{The key difference between Karma and OpenGarden is that 
in the former case, an operator assists the UPN service (by addressing the security and incentive issues), while in the latter case, mobile users operate autonomously without relying on an operator.  
In this work, we will study the \emph{operator-assisted  mobile UPN model} motivated by Karma \cite{Karma}, and focus on both the user's membership selection mechanism and the operation's pricing/incentive design issue.}

The success of a UPN service   relies on the users' willingness to contribute their Internet connectivities and network resources. 
Hence, it is important to design effective \emph{incentive} mechanisms for encouraging mobile users to serve as hosts \cite{UPN-magazine}. 
As an example, the Karma operator offers some free data quota reward to those hosts who share their 4G Internet connections \cite{Karma}. 
Several recent studies further proposed improved incentive mechanisms for UPN \cite{karma-lin, opengarden-lin}, motivated by the commercial practices. 
\revhh{Specifically, in \cite{karma-lin}, Gao \emph{et al.} considered a Karma-like UPN model, and studied the operator's problem of designing the best pricing and free data quota rewarding strategy for a single host, considering the data demands from both the host and his clients.} 
In \cite{opengarden-lin}, Iosifidis \emph{et al.} considered an OpenGarden-like UPN network, and designed a distributed bargaining-based virtual currency system that incentivizes nearby mobile users to fairly and efficiently share connectivity and resources. 
However, these prior studies either focus on the interaction between a particular host and his clients \cite{karma-lin}, or on a static network with a fixed topology \cite{opengarden-lin}, and  
none of them considers issues encountered in a mobile network without a fixed topology, and how the users' membership choices evolve in such networks. 
Starting from some stylized network models, this paper represents a first step towards understanding these important practical network design and operational issues. 

\subsection{Model and Contributions}

In this work, we consider an \emph{operator-assisted  mobile UPN} facilitated by an MVNO. 
\revhh{Similar to Karma, the MVNO is a special wireless service provider, who does not own radio spectrum and/or wireless network infrastructure, but lease these resources from traditional Mobile Network Operators (MNOs) \cite{mono}. 
Comparing with traditional MNOs, an MVNO can employ novel and flexible services and business
models better tailored to the needs of their subscribers, hence can better reach   niche markets that are  under-served by   MNOs.} 
In our model, the MVNO offers Internet access service to two types of subscribers (\emph{hosts} and \emph{clients}) based on the UPN: 
\begin{itemize}
\item 
A \emph{host} can connect to the MVNO's network directly (via a customized portable device) and pay a usage-based price. 
He can also operate as a mobile hotspot and share his Internet connection with clients, and get some free data quota reward from the MVNO as the incentive. 
\item
A \emph{client} can only connect to the MVNO's network via a nearby host and pay a usage-based price. 
\end{itemize}

Given these two choices, each mobile user needs to decide whether to subscribe to the MVNO, and if so, which membership (host or client) to choose. 
A mobile user can also choose not to subscribe to the MVNO and become an \emph{alien}.

\begin{figure}[t]
\vspace{-5mm}
\centering
\includegraphics[width=3in]{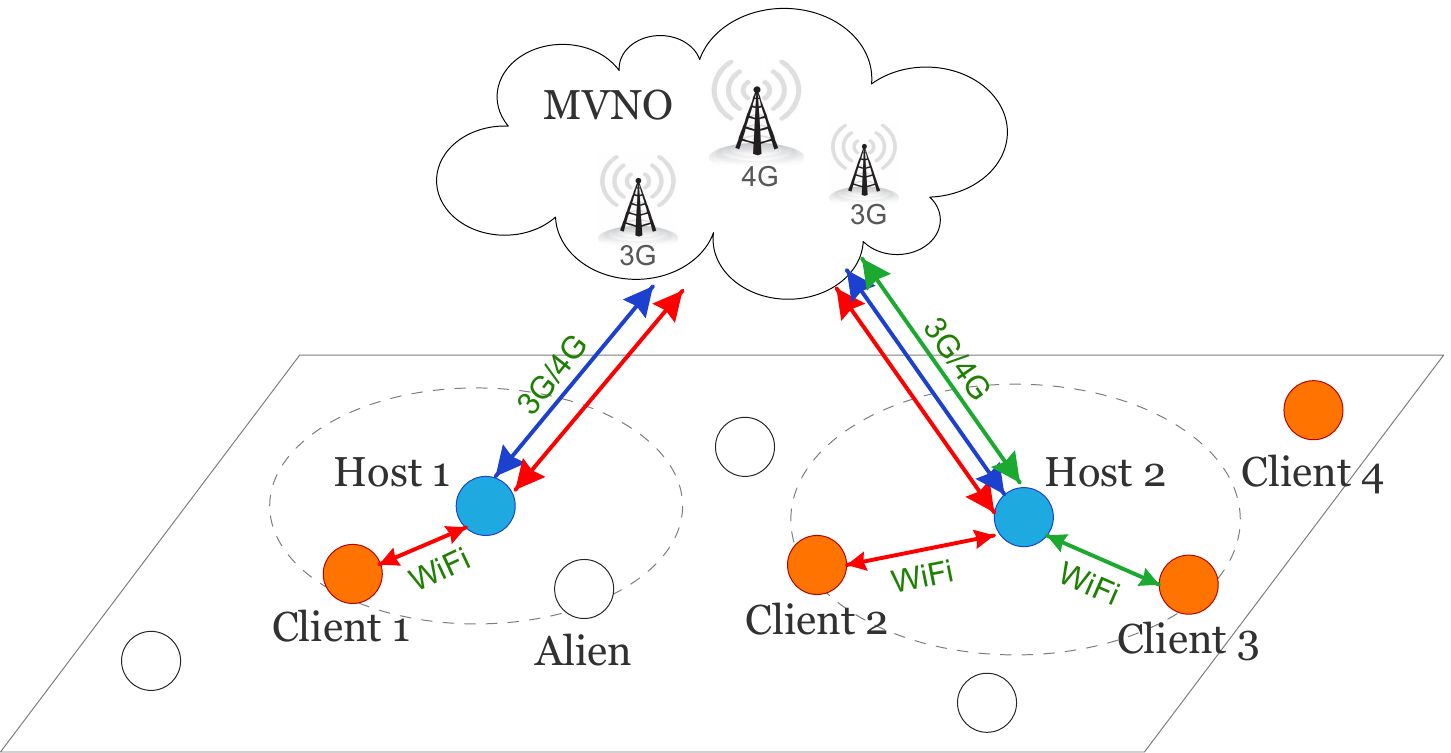}
 \DeclareGraphicsExtensions.
\caption{System Model: Hosts (Blue), Clients (Orange), and Aliens (White).}
\label{fig:model}
\vspace{-5mm}
\end{figure}

Figure \ref{fig:model} illustrates such an operator-assisted  mobile UPN model, where hosts, clients, and aliens are denoted by blue, orange, and white circles, respectively. In this example, client 1 connects to the MVNO's network through host 1, clients 2 and 3 connect to the MVNO's network through host 2, and client 4 cannot connect to the MVNO's network at this moment, as no nearby host can serve it at this instant. 
As users move around in the network, client 4 may eventually meet some hosts  through which he connect to the Internet later on.   

In such an operator-assisted  UPN model, we are interested in answering two key questions: 
\begin{enumerate} 
\item
What is the best membership choice for each mobile user, and how different users' choices affect each other? 
\item 
What is the MVNO's best pricing and free data quota rewarding strategy that maximizes her profit? 
\end{enumerate}
We study the above problems from the game-theoretic perspective. More specifically, we formulate the interactions among the MVNO and mobile users  as a two-stage Stackelberg game, and analyze the game equilibrium systematically. 

In Stage I, the MVNO  determines the best price and free data quota reward to maximize her profit gained from all subscribers (hosts and clients). 
\revhh{A higher price will attract less subscribers, but can generate a higher revenue per subscriber. } 
\revhh{A higher free data quota reward (to hosts) makes it more attractive to be a host, hence will increase the number of hosts. 
Consequently, as the number of hosts increases, clients can find hosts more easily, hence more clients will have incentive to joint the UPN. 
On the other hand, a higher free data quota will naturally reduce the revenue that the MVNO can achieve from each client (due to a higher reward to host).}

In Stage II, mobile users select their best membership types (i.e., host, client, or alien), given the price and free data quota reward provided by the MVNO. 
Note that different users' decisions are coupled with each other. 
For example,  the payoff of a client increases as the number of hosts in the network increases, due to an increased chance of accessing the Internet.  
Similarly, a host's payoff increases as the number of clients increases, but its payoff is reduced as the number of hosts is increased due to the increased competition among hosts.  
We  characterize how different users' membership selections affect each other, and how the membership distribution  dynamically evolves over time  until reaching a stable distribution (called \emph{membership selection equilibrium}).~~~~


The key contributions are summarized as follows. 

\begin{itemize}

\item
This is the first work that systemically studies and characterizes the membership evolution and the corresponding equilibrium in a mobile UPN. 

\item 
Based on the membership selection equilibrium analysis, we further characterize the MVNO's best pricing and incentive mechanism that maximizes her profit. 

\item 
\revx{We perform extensive numerical studies and show that the MVNO's maximum profit increases with the user density under the proposed hybrid pricing and incentive mechanism, and the profit gain can be up to 50\% in a dense network comparing with the optimized pricing-only benchmark without incentives.}

\end{itemize}

The rest of the paper is organized as follows. In Section
\ref{sec:model}, we present the system model. In Sections \ref{sec:analysis}, we analyze the proposed two-stage game. 
In Section \ref{sec:simulation}, we present the simulations and finally, Section \ref{sec:conclusion} concludes the paper.

\section{System Model}\label{sec:model}

\subsection{The Detailed Network Model}
\label{sec:model-network}


We consider a UPN with one MVNO and a set $\N =\lbrace 1,2,\cdots ,N\rbrace$ of mobile users. 
Similar to Karma \cite{Karma}, the MVNO provides   Internet access  to its   subscribers using wireless resources leased from traditional MNOs.
Each mobile user decides whether to subscribe to the MVNO, and if so, what membership type (host or client) to choose. 
Users' decisions not only depend on the MVNO's pricing and incentive mechanism (to be defined in Section \ref{sec:model-user}), but also on how often they meet each other (and hence can form a UPN).~~~~~~~~~~~ 

We consider a time-slotted system, where mobile users may change their locations and hence have different neighbours in different time slots.  
As a first step to understand the impact of \emph{mobility} on the users' choices, we consider a simple model based on the Erdos-Renyi random graph  model widely-used in graph theory \cite{mobility}: Mobile users move randomly in a certain area, and each user has the same probability $\rho \in (0,1)$ of meeting any other user in each time slot.\footnote{Here ``meeting'' means that they are close enough and can form a UPN.}  
This meeting probability does not change over time, and the meeting events are independent across time slots.
\revx{Such a homogeneous mobility model allows us to get closed form expressions and gain valuable insights into the design of UPN. 
We will consider more detailed mobility models in our future work}.~~~~~~~~~~~~~~~~~~~

For analytical convenience, we  assume a large network with   many users (e.g., $N \rightarrow \infty$) and small meeting probability $\rho$ (e.g., $\rho \rightarrow 0 $). 
\revhh{This assumption is mainly used for obtaining the closed form result, and our analysis method holds for small networks as well.} 
To make the model meaningful, we assume that the \emph{average} number of neighbours that a user meets in each time slot, denoted by $\lambda \eq N \cdot \rho $, is a   finite value.


\vspace{-2mm}

\subsection{The MVNO Model}\label{sec:model-mvno}

\rev{The MVNO's goal is to maximize her profit, which depends on the revenue collected from her subscribers (hosts and clients).} 
Specifically, the MVNO charges her subscribers a \emph{usage-based price} for the data consumption, while offering a \emph{free data quota} to encourage a host to provide Internet access for his clients. 
This quota is proportional to the total amount of data that the host forwards for clients. 

Let $p \in [0, p_{\textsc{max}}]$ denote the price per byte of data usage (for both  hosts and clients), and $\delta \in [0, 1]$ denote the free data quota ratio (for hosts). 
Namely, if a host forwards $D$ bytes of data for clients, he can get $\delta D$ bytes of free data from the MVNO.
Clearly, a larger $\delta$ encourages more users to subscribe as hosts.  
The objective of the MVNO is to determine the proper $p $ and $ \delta  $ to maximize her profit.

It should be noted that with the free data quota, two nearby hosts may pretend to be clients and each connects to the other's MiFi device. 
By doing so, both hosts gain certain free data quota by forwarding data for the other user, and both users are better off with no benefit to the network. 
To prevent such an arbitrage  due to collusion, the MVNO offers a \emph{price discount} equal to the free data quota ratio $ \delta $  for the data usage of each host, i.e., a host only pays a price of $p(1-\delta)$ for each unit of his own data.\footnote{\rev{Our analysis can be applied to a more general case with \revhh{a host price (for all hosts) and a client price (for all clients).} 
Here we use the same price (for hosts and clients) in order to comply with the real Karma case \cite{Karma}.}} 
Obviously, with this modification, a host has no incentive to collude with other and  pretend to be a client.~~~~


 \subsection{The Mobile User Model} \label{sec:model-user}

Each user has a service request probability $\theta \in [0,1]$, referred to as the \emph{type} of user. 
Different users may have different types, which are independent and identically distributed (i.i.d) according to a probability distribution function $f(\theta)$. 
Such a service request probability can also represent  the user's  QoS requirement. For example, a user with a higher  data rate  requirement will have a larger $\theta$. 

\revx{In this work, we focus on the \emph{symmetric equilibrium} where users with the same type $\theta$ will always make the same membership choice. 
Moreover, we focus on the user \emph{pure-strategy behaviour} where each user will choose a specific membership under a given network situation.}  
Namely, a user makes his membership choice $s\in \lbrace \h,\c,\a\rbrace $, where
\begin{itemize} 
  \item $\h$: Subscribe to the MVNO as a \emph{host}; 
  \item $\c$: Subscribe to the MVNO as a \emph{client}; 
  \item $\a$: Do not subscribe to the MVNO and remain as an \emph{alien}.  
\end{itemize}   

\revx{The \emph{payoff} of each user is the difference between the achieved benefit and the incurred cost.} 
For convenience, we denote the payoff of a type-$\theta$ user under  membership $s$ by $U_{\theta}(s)$. 
The objective of the user is to choose the proper membership to maximize his payoff. 

\emph{1) Alien:} 
An alien does not subscribe to the MVNO, hence cannot connect to the MVNO's network.
He may access the Internet through other means. 
Without loss of generality, we normalize an alien's payoff to be zero:\footnote{If an alien's payoff is a positive constant, we can always normalize it to zero by redefining the cost terms of $\Ch$ and $\Cc$ later on.} 
\begin{equation}
U_{\theta}( \a)=0.
\end{equation}

\emph{2) Client:} 
A client can only connect to the MVNO's network via an MiFi hotspot provided by a nearby host. 
Hence, the payoff of a client   depends on the probability that he can meet hosts. 
If a client meets multiple hosts at the same time, \emph{he   randomly chooses one host  for his   connection}.  
Let $P_{\h}$ denote the probability of a client meeting at least one host in a time slot (derived in Section \ref{sec:analysis-0}). 
Then, the expected payoff (per time slot) of a type-$\theta$ client is 
\begin{equation}
\begin{aligned}
U_{\theta}( \c) = & \ \theta \cdot P_{\h} \cdot \underbrace{(\bar{v}_{\c} - \costc  - p)}_{\Pi_\c}- \Cc
,
\end{aligned}
\end{equation}
where $\bar{v}_{\c}$ is the average data value for a client, $\costc$ is the average transmission cost of a client  (e.g., the energy cost), and $\Cc$ is the time-average cost of being a client (e.g., the subscription fee and advance payment). 
For notational convenience, we denote $\Pi_\c = \bar{v}_{\c} - \costc  - p$ as the client's average benefit of consuming one byte of data.
 
\emph{3) Host:} 
A host can connect to the MVNO's network anytime via his MiFi device. 
Meanwhile, this MiFi device can also provide Internet connections for nearby clients, and  the host can gain certain free data quota (depending on the total amount of data that he forwards for clients). 
It is important to note that \emph{a host shares only his Internet connection, but not his data, with clients}. That is, clients need to pay for their own data usage. 
Clearly, the payoff of a host depends on the number of clients that he serves as well as the data usage of his served clients. 
Let $Y_{\c}$ denote the average number of clients that a host serves in a time slot (derived in Section \ref{sec:analysis-0}), and $\bar{\theta}_{\c}$ denote the average service request probability of   clients (derived in Section \ref{sec:analysis-0}). Then, the expected payoff (per time slot) of a type-$\theta$ host is\footnote{\revx{This equation implies that if the rewarded free data quota is larger than a host's actual data request, he can either transfer the excessive free data into the monetary reward, or reserve the excessive free data for the future usage.}}
\begin{equation}
\begin{aligned}
&  U_{\theta}( \h)= 
\\
& \theta \cdot  \underbrace{\left( \bar{v}_{\h} - \costh - p   (1-\delta) \right)}_{\Pi_\h}  + \ \bar{\theta}_{\c} \cdot   Y_{\c} \cdot    \underbrace{(\delta   p- \costhc)}_{\widetilde{\Pi}_\h} -\Ch , &
\end{aligned}
\end{equation}
where $\bar{v}_{\h}$ is the average data value for a host, $\costh $ and $ \costhc$ are the average transmission costs of a host for his own data and for his served clients' data, respectively, 
$ \Ch$ is the time-average cost of being a host (e.g., the cost of purchasing the MiFi device, the subscription fee, and advance payment). 
For notational convenience, we denote $\Pi_\h = \bar{v}_{\h} - \costh - p\cdot (1-\delta)$ as the host's average benefit of consuming one byte of data, and 
$\widetilde{\Pi}_\h = \delta\cdot p- \costhc$ as the host's average benefit of forwarding one byte of data for his clients. 

To avoid the trivial scenarios, we further assume:  
\begin{itemize}
\item $\Pi_\h > \Pi_\c$: The host's average data usage benefit is larger than that of a client.    
This is   because the host can   reserve higher quality wireless resources  for his own transmission, and   the data price for the host is discounted;  
\item $\Ch > \Cc$: The time-average cost of being a  host is higher than that of a client. 
This is   because the host needs to purchase and maintain the MiFi device;
\item $\widetilde{\Pi}_\h > 0 $: That is, hosts can always benefit from serving clients. 
This is important for encouraging hosts to share their Internet connections with clients. 
If this is violated, then there will be no UPN. 
\end{itemize}

\subsection{Problem Definition}

We formulate the interactions between the MVNO and mobile users as a two-stage Stackelberg game. 

In Stage I, the MVNO decides the price $p$ and the free data quota ratio $\delta$  to maximize the expected profit.
Notice that the MVNO earns $p\cdot (1-\delta)$ per byte of data from both hosts and clients, and meanwhile pays the MNOs  a resource leasing cost   $\omega$ for his subscribers' data consumption.  
Thus, the expected profit (per time slot) of the MVNO is 
\begin{equation}\label{eq:vp}
V(p,\delta)=
 ( \mu_\h \bar{\theta}_{\h} + P_{\h} \cdot \mu_\c \bar{\theta}_{\c}) \cdot
(p\cdot (1-\delta)-\omega) \cdot N, 
\end{equation}  
where 
$\mu_{\h}$ and $\mu_{\c}$ denote the percentages of hosts and clients, respectively, $\bar{\theta}_{\h}$ and $\bar{\theta}_{\c}$ denote the average service request probabilities of hosts and clients, respectively.  
The scale factor $ P_{\h} $ on $  \bar{\theta}_{\c} $ implies that a client can consume data only when he meets at least one host.
Keep in mind that $\mu_\h, \mu_\c , \bar{\theta}_{\h}, \bar{\theta}_{\c}$, and $P_{\h}$ are all functions of $(p,\delta)$ derived in Stage II. 


In Stage II, mobile users decide their memberships (i.e., host, client, or alien), given the price $p$ and free data quota ratio $\delta$ announced by the MVNO in Stage I. 
Different users' decisions are coupled. 
For example, with more users choosing to be hosts, the payoff of being a client will  increase (due to a larger meeting probability  $P_{\h}$). Similarly, with more users choosing to be clients, the payoff of being a host will   increase (due to a larger client number $Y_{\c}$).


\section{Game Equilibrium Analysis}\label{sec:analysis}

In this section, we will study the Subgame Perfect Equilibrium (SPE) of the proposed two-stage Stackelberg game. 

\begin{definition}[Subgame Perfect Equilibrium (SPE)]
A strategy profile $\{(p^*, \delta^*),$ $ (s^*(\theta),\forall \theta \in [0, 1])\}$, where $(p^*, \delta^*)$ is the MVNO's strategy in Stage I and $s^*(\theta)$ is a type-$\theta$ user's strategy in Stage II, is an SPE if and only if \footnote{To be more precise, Stage II equilibrium should be written as $s^*(\theta,p,\delta)$, as the membership selection is a function of the type $\theta$, price $p$, and free data quota $\delta$. We will keep $s^*(\theta)$ for the simplicity of presentation.}
\begin{equation*}
\left\{
\begin{aligned}
\mbox{Stage II:~} & U_{\theta} (s^*(\theta) ) \geq U_{\theta} (s),\quad \forall \theta\in[0, 1], s\in\{\a,\c,\h\};
\\
\mbox{Stage I:~} & V(p^*,\delta^*) \geq  V(p ,\delta ),\quad \forall 
p \in [0, p_{\textsc{max}}], \delta \in [0, 1].
\end{aligned}
\right.
\end{equation*}
\end{definition}

We will derive the SPE by backward induction. 
Namely, we first study the user membership selection game in Stage II, and characterize the users' equilibrium membership decisions. 
Then we characterize the MVNO's optimal price and free data quota ratio that maximize her profit in Stage I.   


\subsection{Derivation of Several Important Variables}\label{sec:analysis-0}

Before studying the game equilibrium, we first derive the values of $ P_{\h}$, $Y_{\c}$, $ \bar{\theta}_{\c}$, and $\bar{\theta}_{\h}$ analytically.  

\subsubsection{\textbf{Calculation of $\bar{\theta}_{\c} $ and $\bar{\theta}_{\h} $} (Average service request probabilities of clients and hosts)} \label{sec:analysis-0-x}
For convenience, we denote  ${\Theta}_{\a} $, ${\Theta}_{\c} $, and ${\Theta}_{\h} $ as the sets of service request probability $\theta$ with which users  choose to be aliens, clients, and hosts, respectively. 
\rev{Obviously, ${\Theta}_{\a} $, ${\Theta}_{\c} $, and ${\Theta}_{\h} $ form a
\emph{partition} of the whole user-type set, i.e., ${\Theta}_{\a} \cup {\Theta}_{\c} \cup {\Theta}_{\h} = [0,1]$.} 
Let $\mu_{\a}$, $\mu_{\c}$, and $\mu_{\h}$ denote the \emph{percentages of aliens, clients, and hosts}, respectively. That is, 
\begin{equation}
\mu_{s} = \int_{\theta \in {\Theta}_{s}} f(\theta) \mathrm{d} \theta, \quad \forall s\in\{\a,\c,\h\}. 
\end{equation}
Then, the average service request probabilities of clients and hosts are, respectively,
\begin{equation}\label{eq:average}
\bar{\theta}_{\c} = \frac{1}{\mu_{\c}} \int_{\theta \in {\Theta}_{\c}} \theta \cdot f(\theta) \mathrm{d} \theta ,
\mbox{~~and~~}
\bar{\theta}_{\h} = \frac{1 }{\mu_{\h}} \int_{\theta \in {\Theta}_{\h}} \theta \cdot f(\theta) \mathrm{d} \theta. 
\end{equation}
To obtain closed form results, in the rest of the  analysis, \emph{we assume a uniform distribution for $\theta$, i.e., $f(\theta) = 1, \forall \theta\in[0, 1]$.} 
Note, however, that our analysis method holds for an arbitrary distribution function  $f(\theta)$.

\subsubsection{\textbf{Calculation of $P_{\h} $} (Probability of meeting at least one host)}

Consider a user $i$ in the network. 
Based on our assumption in Section \ref{sec:model-network}, the probability that   user $i$ meets another user $j$ is $\rho$. 
Based on the discussion in Section \ref{sec:analysis-0}.1, the probability that such a user $j$ being a host is $\mu_{\h}$. 
Since there are a total of $N-1$ other users (excluding   $i$), the probability that user $i$ does not meet any host is $ (1 - \rho \mu_{\h})^{N-1}$. 
Thus, the probability that user $i$ meets at least one host is: 
\begin{equation}
P_{\h} =1-(1-\rho \cdot \mu_{\h})^{N-1} . 
\end{equation} 

\rev{As we have assumed that $N$ is very large and $\rho$ is very small,} we can rewrite $P_{\h}$ in the following way:
\begin{equation}
\begin{aligned}	
P_{\h} =  \lim_{N\rightarrow\infty} 1- \left( 1-  { \textstyle \frac{\lambda}{N} \cdot \mu_{\h}} \right)^{ N-1 }
=
 1-e^{-\mu_\h\lambda}, 
\end{aligned}	
\end{equation}
\rev{where $ \lambda = N \cdot\rho $ denotes the average number of users (host, client, or alien) that a user meets in each time slot.}

\subsubsection{\textbf{Calculation of $Y_{\c} $} (Average number of clients connected to a host)} 

Consider an arbitrary host $i$ in this network. 
We first notice that when a client meets multiple hosts at the same time, he will randomly pick one to connect.  
Hence, the probability that host $i$ is chosen by a   client $j$ that he meets is: 
\begin{equation}
P_{\c}^{\textsc{con}} = 1 \cdot P_{\h}^{(0)} + \frac{1}{2} \cdot P_{\h}^{(1)} + ... + \frac{1}{N-1} \cdot P_{\h}^{(N-2)}, 
\end{equation}
where $P_{\h}^{(k)}$ denotes the probability that client $j$ meets $k$ \emph{other} hosts except host $i$. 
We further notice that (i) the probability that client $j$ meets another user is $\rho$, (ii) the probability that such a user being a host is $\mu_{\h}$, and (iii) there are a total of $N-2$ other users (excluding $i$ and $j$) that client $j$ can meet. Hence, the probability that client $j$ meets $k$ \emph{other} hosts is: 
\begin{equation}
P_{\h}^{(k)} = {\textstyle \binom{N-2}{k}} \cdot (\rho\mu_\h)^k\cdot (1-\rho\mu_\h)^{N-2-k}, 
\end{equation}
which follows the binomial distribution (with a total of $N-2$ trials and a success probability $\rho\mu_h$ in each trial). 
As $N$ is very large and $\rho = \frac{\lambda}{N} $ is very small, we will have: 
\begin{equation}
\begin{aligned}
P_{\h}^{(k)} = & \lim_{N\rightarrow\infty} {\textstyle \binom{N-2}{k}}  \cdot 
( { \textstyle \frac{\lambda}{N}   \mu_{\h}} )^k\cdot 
(1- { \textstyle \frac{\lambda}{N} \mu_{\h}})^{N-2- k}
\\
= &\  \frac{(\mu_\h\lambda)^k }{k !} \cdot e^{-\mu_\h\lambda}   ,  
\end{aligned}
\end{equation}
which follows the Poisson distribution (with an arrival rate of $\mu_\h\lambda$). Hence, the probability that host $i$ is chosen by a   client $j$ that he meets can be rewritten as: 
\begin{equation}
\begin{aligned}
P_{\c}^{\textsc{con}} = & \lim_{N\rightarrow\infty} \sum_{k=0}^{N-2} \frac{1}{k+1} \cdot \frac{(\mu_\h\lambda)^k }{k !} \cdot e^{-\mu_\h\lambda} 
\\
= & \ \frac{1}{\mu_\h\lambda} \cdot (1 - e^{-\mu_\h\lambda} ). 
\end{aligned}
\end{equation}

We further notice that the average number of clients that a host $i$ meets in each time slot is $(N-1)  \cdot \rho \mu_\c$. 
Hence, the average number of clients connected to host $i$ is: 
\begin{equation}
Y_{\c} =\lim_{N\rightarrow\infty} (N-1)  \cdot \rho  \mu_\c \cdot P_{\c}^{\textsc{con}}  = \frac{\mu_\c}{\mu_\h}\cdot (1 - e^{-\mu_\h\lambda} ).
\end{equation}

\subsection{Stage II: User Membership Selection Game}\label{sec:analysis-2} 

Now we study the user membership selection game in Stage II, given the MVNO's decision $(p, \delta)$ in Stage I. 

As discussed previously, the sets of aliens, clients, and hosts already in the system (i.e., $\Theta_\a$, $\Theta_\c$, and $\Theta_s$) and their corresponding percentages (i.e.,  $\mu_\a$, $\mu_\c$, and $\mu_s$) will affect the values of $ P_{\h}$, $Y_{\c}$, $ \bar{\theta}_{\c}$, and $\bar{\theta}_{\h}$, and further affect the user payoff and membership selection. 
Hence, in what follows, we will first study what is the user's best membership decision under a particular membership distribution $\{\Theta_\a,\Theta_\c,\Theta_\h\}$. 
Then, we will study how the user membership decision dynamically evolves over time, and what is the stable membership distribution (called  \emph{membership selection equilibrium}).  

\subsubsection{\textbf{User Best Membership Selection}}

Given the MVNO's strategy $(p, \delta)$, and under a particular initial membership distribution $\{\Theta_\a,\Theta_\c,\Theta_\h\}$, the payoff of a type-$\theta$ user is:
\begin{equation}
U_{\theta} (s) = \left\{
\begin{aligned}
& 0 , & \quad \mbox{if } s=\a,
\\
& \theta P_{\h}   \Pi_\c- \Cc, & \quad \mbox{if } s=\c,
\\
& \theta \Pi_\h, + \bar{\theta}_{\c}    Y_{\c}   \tilde{\Pi}_{\h}  -\Ch ,  & \quad \mbox{if } s=\h. 
\end{aligned}
\right.
\end{equation}
A type-$\theta$ user will choose to be a client, if and only if
\begin{equation}\label{eq:uthc}
U_{\theta} (s = \c) \geq \max\{0,\ U_{\theta} (s = \h ) \}, 
\end{equation}
and choose to be a host, if and only if
\begin{equation}\label{eq:uthh}
U_{\theta} (s = \h) \geq \max\{0,\ U_{\theta} (s = \c ) \}. 
\end{equation} 

We further notice that $\Pi_\h  > P_\h \Pi_\c $ (as $\Pi_\h > \Pi_\c$ and $P_\h \leq 1$). 
Then, we can immediately obtain the following observation: (i) if a type-$\theta$ prefers to be a host, then all users with a type larger than $\theta$ prefer to  be a host; and (ii) if a  type-$\theta$ prefers to be an alien, then all users with a type smaller than $\theta$ prefer to be an alien.
Intuitively, this is because a higher type user has a stronger preference of  being a host due to the higher data usage benefit, and a lower type user has a stronger preference of being an alien due to the lower data usage benefit. 


Based on the above, we have the following lemma.\footnote{Due to space limit, we put the proofs in the online technical report \cite{report}.}

\begin{lemma}
The best user membership selection  under a given initial membership distribution $\{\Theta_\a,\Theta_\c,\Theta_\h\}$ can be characterized by two threshold types $\underline{\theta}_{\a }$ and $\underline{\theta}_{ \h}$, where
\begin{equation}\label{eq:threshold}
\begin{aligned}
\underline{\theta}_{\a } = & \ \textstyle \left[ \min \left\{ \frac{\Cc}{P_\h \Pi_\c},\  \frac{ \Ch - \bar{\theta}_{\c}    Y_{\c}   \tilde{\Pi}_{\h} }{\Pi_\h}  \right\} \right]_{0}^{1},
\\
\underline{\theta}_{\h} = & \ \textstyle \left[   \max \left\{  \frac{ \Ch - \bar{\theta}_{\c}    Y_{\c}   \tilde{\Pi}_{\h} }{\Pi_\h},\  \frac{\Ch-\Cc - \bar{\theta}_{\c}    Y_{\c}   \tilde{\Pi}_{\h}}{\Pi_\h - P_\h \Pi_\c} \right\} \right]_0^1 , 
\end{aligned}
\end{equation}
with $[x]_0^1  \eq \min \{ \max\{ x, 0\}, 1 \}$.  
Here, $\underline{\theta}_{\a } $ denotes the largest $\theta$   that a type-$\theta$ user prefers to be an alien, and $\underline{\theta}_{\h}$ denotes the smallest $\theta$   that a  type-$\theta$ user prefers to be a host. 
Moreover, users with a type between $\underline{\theta}_{\a }$ and $\underline{\theta}_{\h }$ prefer to be clients.
\end{lemma}

\revhh{It is easy to see that $\underline{\theta}_{\a } \leq  \underline{\theta}_{\h }$, as $\underline{\theta}_{\a }  \leq [ \frac{ \Ch - \bar{\theta}_{\c}    Y_{\c}   \tilde{\Pi}_{\h} }{\Pi_\h} ]_0^1   \leq \underline{\theta}_{\h } $ according to \eqref{eq:threshold}.
Note that  $\underline{\theta}_{\a } = \underline{\theta}_{\h }  $ implies that  no user chooses to be a client.}

Obviously, under the user best membership selection, the newly \emph{derived} membership distribution may be different from the initial one. We denote the newly derived membership distribution by $\{\Theta_\a',\Theta_\c',\Theta_\h'\}$, and the associated membership  percentages as $\{\mu_\a',\mu_\c',\mu_\h'\}$. Formally, 
\begin{proposition}
Given an initial distribution $\{\Theta_\a,\Theta_\c,\Theta_\h\}$, the newly derived membership distribution $\{\Theta_\a',\Theta_\c',\Theta_\h'\}$ is 
\begin{equation}\label{eq:derived-dis}
 \Theta_\a' = [0, \underline{\theta}_{\a }],\ \ 
 \Theta_\c' = [\underline{\theta}_{\a }, \underline{\theta}_{\h}],\  \ 
 \Theta_\h' = [\underline{\theta}_{\h }, 1], 
\end{equation}
and the associated percentages $\{\mu_\a',\mu_\c',\mu_\h'\}$ are given by: 
\begin{equation}\label{eq:derived-per}
\mu_\a' = \underline{\theta}_{\a }, \ \ 
\mu_\c' = \underline{\theta}_{\h } - \underline{\theta}_{\a },\ \ 
\mu_\h' = 1 -  \underline{\theta}_{\h }.
\end{equation}
\end{proposition}

Note that in the above analysis, we assume that all users make simultaneous best membership selections.

\subsubsection{\textbf{Membership Distribution Dynamics}}


The membership distribution change in (\ref{eq:derived-dis}) and (\ref{eq:derived-per}) continues evolving over time, 
until it reaches a stable distribution (called \emph{membership selection  equilibrium}), where no user has the incentive to change his choice. 
Now we   study such a
membership distribution dynamics, and characterize the membership selection equilibrium in Stage II. 

Suppose that each user selects membership once in each time slot. 
Without loss of generality, we consider the membership distribution change in a generic time slot $t$.
Let $\{\Theta_\a^t,\Theta_\c^t,\Theta_\h^t\}$ denote the initial membership distribution at the beginning of slot $t$, and $\{\Theta_\a^{t+1},\Theta_\c^{t+1},\Theta_\h^{t+1}\}$ denote the newly derived membership distribution in time slot $t$ (after one round best response of each user). 
A \emph{membership selection equilibrium} is characterized by the following proposition. 
\begin{proposition}
A membership distribution $\{\Theta_\a^t,\Theta_\c^t,\Theta_\h^t\}$
is a membership selection equilibrium if
and only if 
$$ 
\Theta_\a^{t+1} = \Theta_\a^t,\ \ 
\Theta_\c^{t+1} = \Theta_\c^t,\ \ 
\Theta_\h^{t+1} = \Theta_\h^t.
$$
\end{proposition}

This implies that if $\{\Theta_\a^t,\Theta_\c^t,\Theta_\h^t\}$ is a   membership selection equilibrium, then we will have: $\Theta_s^{\tau} = \Theta_s^{t}$, $\forall s\in\{\a,\c,\h\}$, for all future time slots $\tau > t$. 
Notice that a membership distribution $\{\Theta_\a,\Theta_\c,\Theta_\h\}$ can be fully characterized by  threshold types $\{\underline{\theta}_{\a }, \underline{\theta}_{\h}\}$ 
or membership percentages $\{\mu_{\c }, \mu_{\h}\}$. 
Hence, we can characterize the membership selection equilibrium by 
$$ 
\mbox{(i)~~~~} \underline{\theta}_{\a }^{t+1} = \underline{\theta}_{\a }^t,\ \ 
\underline{\theta}_{\h}^{t+1} = \underline{\theta}_{\h}^t, 
$$
$$
\mbox{or~~(ii)~~~~} \mu_{\c }^{t+1} = \mu_\c^t,\ \ 
\mu_{\h}^{t+1} = \mu_{\h}^t.~~~
$$
Thus, we have the following lemma for computing the equilibrium with respect to $\{\mu_{\c }, \mu_{\h}\}$. 

\begin{lemma}
The membership selection equilibrium  $\{\mu_{\c }, \mu_{\h}\}$ satisfies the following   equations: 
\begin{equation}\label{eq:networkeq}
\left\{
\begin{aligned}
\mu_{\c } = & \ \underline{\theta}_{\h} (\mu_\c, \mu_\h) -  \underline{\theta}_{\a} (\mu_\c, \mu_\h) , 
\\
\mu_{\h} = & \   1 -\underline{\theta}_{\h} (\mu_\c, \mu_\h), 
\end{aligned}
\right. 
\end{equation}
where $ \underline{\theta}_{\h} (\mu_\c, \mu_\h)  $ and $
\underline{\theta}_{\a } (\mu_\c, \mu_\h) $ are threshold types defined in (\ref{eq:threshold}), and are functions of percentages $\{\mu_{\c }, \mu_{\h}\}$.
\end{lemma}

%
%
%
%

We next discuss the existence and uniqueness of membership selection equilibrium. 

\begin{theorem}[Existence]\label{theo:1}
There exists at least one equilibrium in the membership selection game in Stage II. 
\end{theorem}

However, in general we cannot guarantee the uniqueness of equilibrium. In fact, there may be multiple solutions for the equations in (\ref{eq:networkeq}), which implies that there may be multiple equilibria. 
Nevertheless, 
it can be verified that when multiple equilibria exists, the final equilibrium point is uniquely determined by the initial membership distribution. 
Thus, we have the following conditional uniqueness. 

\begin{theorem}[Uniqueness]
Given any initial membership distribution, the simultaneous best response dynamics evolve to a unique membership selection equilibrium.
\end{theorem}

In this work, we assume an initial distribution with $\mu_\c = 0$ and $ \mu_\h = 0 $, i.e., all users are aliens initially. 
This is a reasonable  assumption, as all users are aliens before joining the UPN. 
We further illustrate the dynamics of membership percentages in \cite{report}.



%
%
%
%

\vspace{-3mm}

\subsection{Stage I: MVNO's Best Decision}\label{sec:analysis-1}

Now we study the MVNO's best price and free data quota  ratio strategy $(p, \delta)$ in Stage I. 

We have shown that given any MVNO strategy $(p, \delta)$, the user membership selection game will converge to a unique equilibrium (from the initial state $\mu_\c = 0$ and $ \mu_\h = 0 $), denoted by  $\{\mu_\c^* , \mu_\h^* \}$ or  $\{\mu_\c^* (p, \delta), \mu_\h^* (p, \delta) \}$.
Equivalently, we can write the equilibrium membership distribution
$\{\Theta_\a^*,\Theta_\c^*,\Theta_\h^*\}$ as 
\begin{equation}
\left\{
\begin{aligned}
\Theta_\a^* = &\ [0,\  1- \mu_\c^*-\mu_\h^*], 
\\
\Theta_\c^* =&\  (1- \mu_\c^*-\mu_\h^*,\  1- \mu_\h^*], 
\\
\Theta_\h^* = &\ (1- \mu_\h^*,\  1]. 
\end{aligned}
\right.
\end{equation}
By Eq. (\ref{eq:average}), we can further derive the average service request probabilities of clients and hosts under the membership selection equilibrium, i.e.,
\begin{equation}
\left\{
\begin{aligned}
\bar{\theta}_\c^* = & \textstyle  \frac{\int_{\theta \in \Theta_\c^* } \theta \cdot f(\theta) \mathrm{d} \theta  }{\mu_\c^*}  = \frac{2 - 2 \mu_\h^* -\mu_\c^*}{2},
\\
\bar{\theta}_\h^* = & \textstyle  \frac{\int_{\theta \in \Theta_\h^* } \theta \cdot f(\theta) \mathrm{d} \theta  }{\mu_\h^*}  = \frac{2 - \mu_\h^*}{2}.
\end{aligned}
\right.
\end{equation}

It is easy to see that $\bar{\theta}_\c^*$ and $\bar{\theta}_\h^*$ are functions of $p$ and $\delta$ (as $ \mu_\c^* $ and $ \mu_\h^* $ are functions of $p$ and $\delta$), and hence can be written as $\bar{\theta}_\c^* (p, \delta)$ and $\bar{\theta}_\h^* (p, \delta)$.
Therefore, by (\ref{eq:vp}), the MVNO's profit optimization problem can be written as 
\begin{equation}\label{eq:mvno-problem}
\max_{ \{p,\delta\} } \ \big(X_\h^*(p, \delta) + X_\c^*(p, \delta) \big) \cdot  \big(p(1-\delta)-\omega \big),
\end{equation}
where $X_\h^*(p, \delta) = N \cdot  \mu_\h^*(p, \delta) \cdot \bar{\theta}_\h^*(p, \delta) $ and $X_\c^*(p, \delta) =  N\cdot  \mu_\c^*(p, \delta) \cdot \bar{\theta}_\c^*  (p, \delta)
\cdot P_\h(p, \delta) $ denote the average amount of data that hosts and clients consume, respectively. 


We can check that the MVNO's optimization problem (\ref{eq:mvno-problem}) is non-convex. Hence,  it is difficult to obtain the closed form solution of the optimal price and free data quota ratio $(p^*,\delta^*)$. 
Fortunately, problem (\ref{eq:mvno-problem}) is a two-variable optimization problem with box constraint sets (i.e., $p \in [0, p_{\textsc{max}}]$ and $\delta \in [0, 1]$), and hence can be effectively solved by using certain numerical methods such as the branch-and-bound method. 

We solve problem (\ref{eq:mvno-problem}) in the following sequential manner. First, we find the optimal price $p^*(\delta)$ under any $\delta$ through one dimensional search. Second, we find the optimal $\delta^*(p)$ under any price $p$ through one dimensional search.  
Then, 

\begin{proposition}
The MVNO's best strategy $(p^*,\delta^*)$ must
occur at an intersection point of $p^*(\delta)$ and $\delta^*(p)$. 
\end{proposition}


\section{Simulation}\label{sec:simulation}

\begin{figure}[t]
 \centering
\includegraphics[width=2.6in]{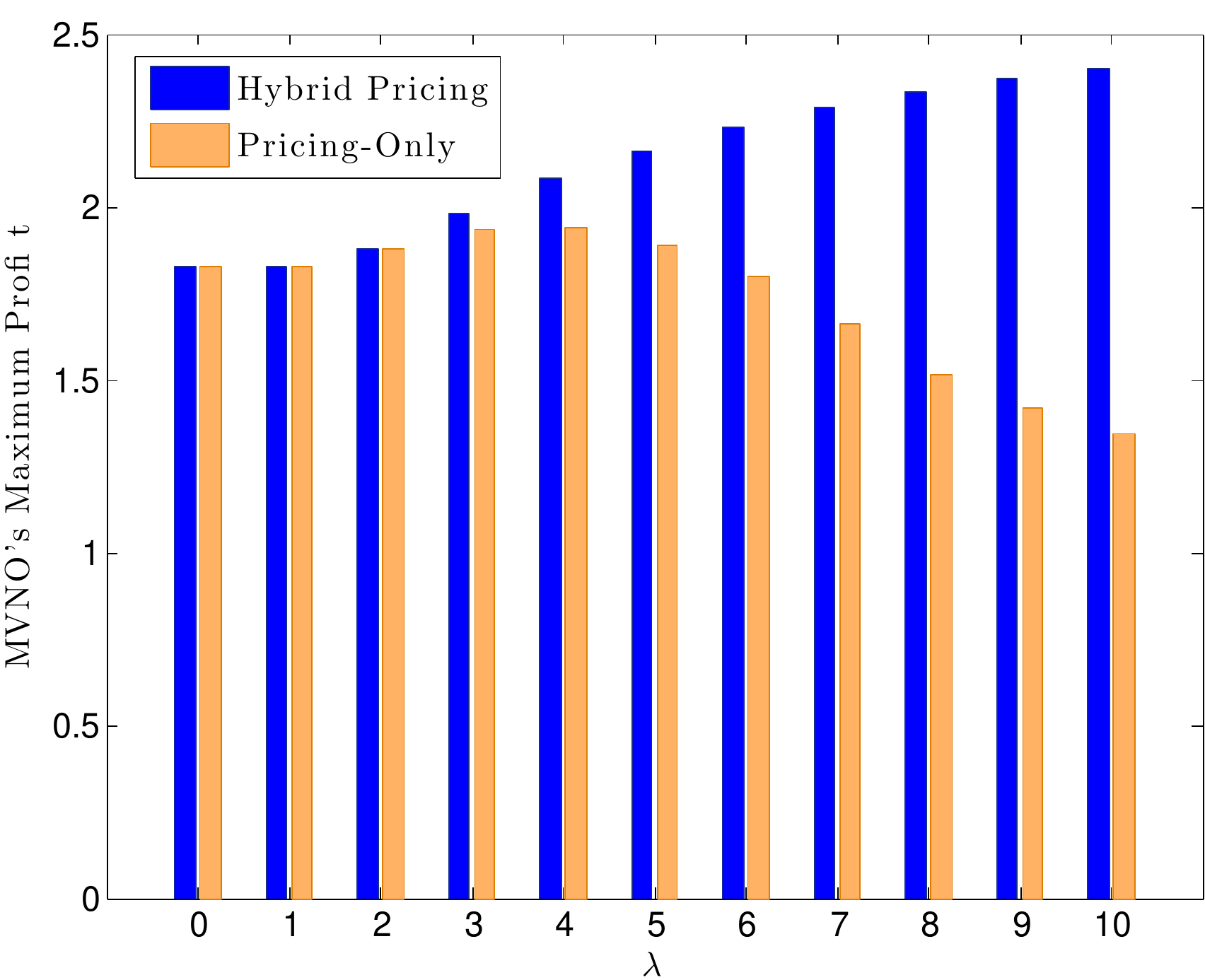}
 \vspace{-3mm}
 \caption{MVNO's maximum profit as a function of $\lambda$}
\label{fig:maxpayoff-a}
 \vspace{-3mm}
\end{figure}

\revx{We perform numerical studies in a network with the following parameters: 
$\bar{v}_{\h}= 15, \bar{v}_{\c}= 10,$ 
$\Ch= 5, \Cc= 1,$ 
$\gamma_\h= 0.5, \gamma_{\h\c}= 1, \gamma_{\c}= 0.1,$ 
and $\omega= 0.5$. 
Notice that we assume that a host's transmission cost for his own data ($\gamma_\h$) is smaller than that for a client's data ($\gamma_{\h\c}$), as the host not only needs to forward the client data to the MVNO's network  via the 3G/4G cellular connection, but also needs to communicate with the client through the Wi-Fi connection. 

In Figure \ref{fig:maxpayoff-a}, we present the MVNO's maximum profits under the proposed best pricing and incentive mechanism (called \emph{hybrid pricing})  
and the optimal \emph{pricing-only} strategy without incentive (i.e., for a fixed $\delta = 0$).
\revhh{Figure \ref{fig:maxpayoff-a} shows that as long as $\lambda>0$, our proposed hybrid pricing strategy always outperforms the pricing-only strategy in terms of the MVNO's profit.
Such a profit gain increases with $\lambda$, and can reach 50\% when $\lambda = 10$.} 
This implies that the MVNO can achieve a larger gain in a dense network with a larger $\lambda$.

Due to space limit, we put more simulation results in the online technical report \cite{report}.}

%

\section{Conclusion}\label{sec:conclusion}

In this work, we first studied the user membership evolution in an  operator-assisted UPN, and then studied the MVNO's best strategy that maximizes her profit. 
Our studies address the incentive and optimality issues in a UPN   from the system perspective. 
There are several interesting directions for future research, such as considering more realistic mobility models and the impact of users' social relationship on the membership selection and Internet connection sharing decisions.

\begin{figure*}[t]
\centering
\begin{minipage}[t]{0.3 \linewidth}
\centering
\includegraphics[height=1.6 in]{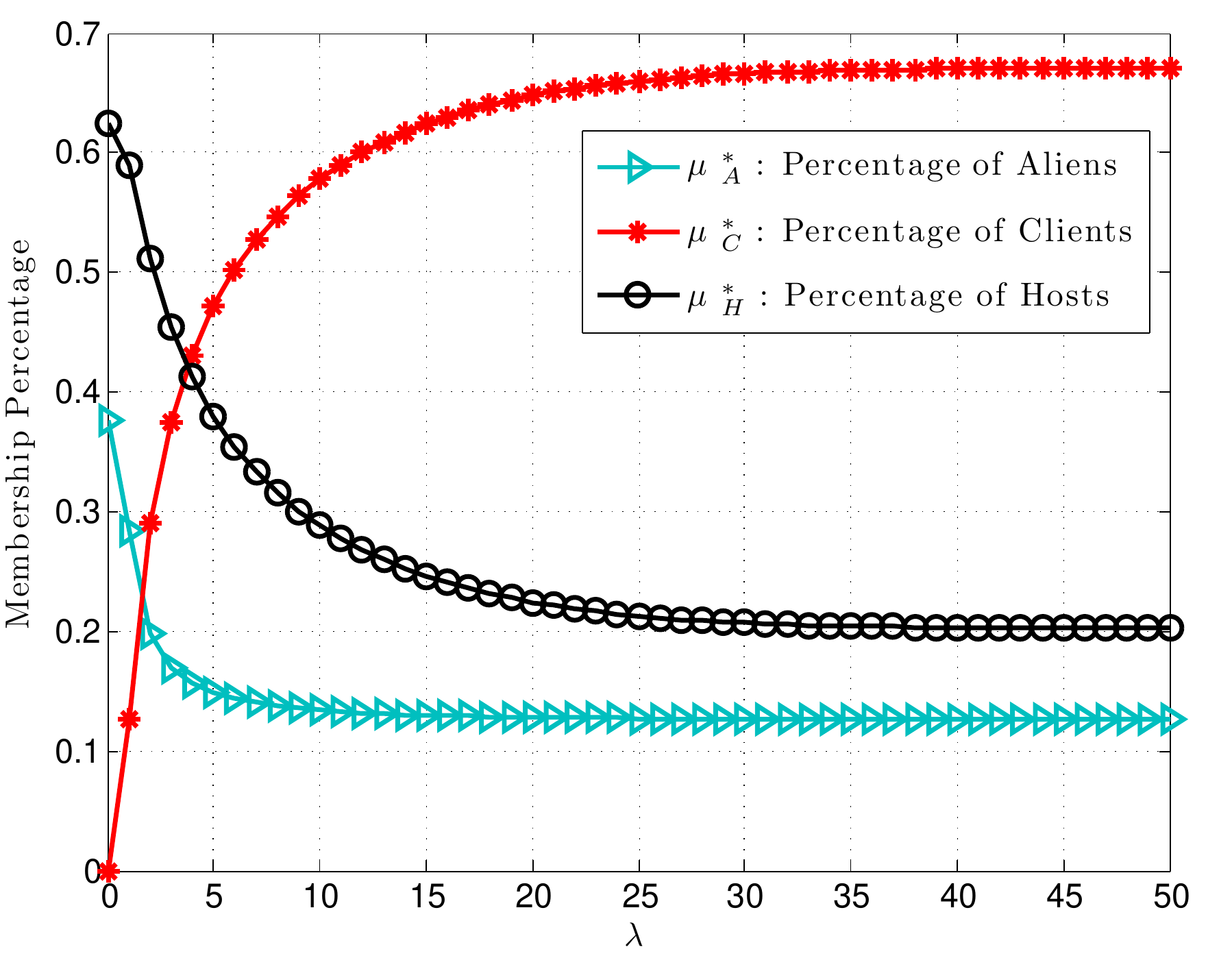}
\caption{  Equilibrium   \{$\mu^*_\a$, $\mu^*_\c$, $\mu^*_\h$\} vs $\lambda$
 (when $p=2$ and $\delta=0.4$)}
\label{fig:lambda}
\end{minipage}
\begin{minipage}[t]{0.03 \linewidth}
~
\end{minipage}
\begin{minipage}[t]{0.3 \linewidth}
\centering
\includegraphics[height=1.6 in]{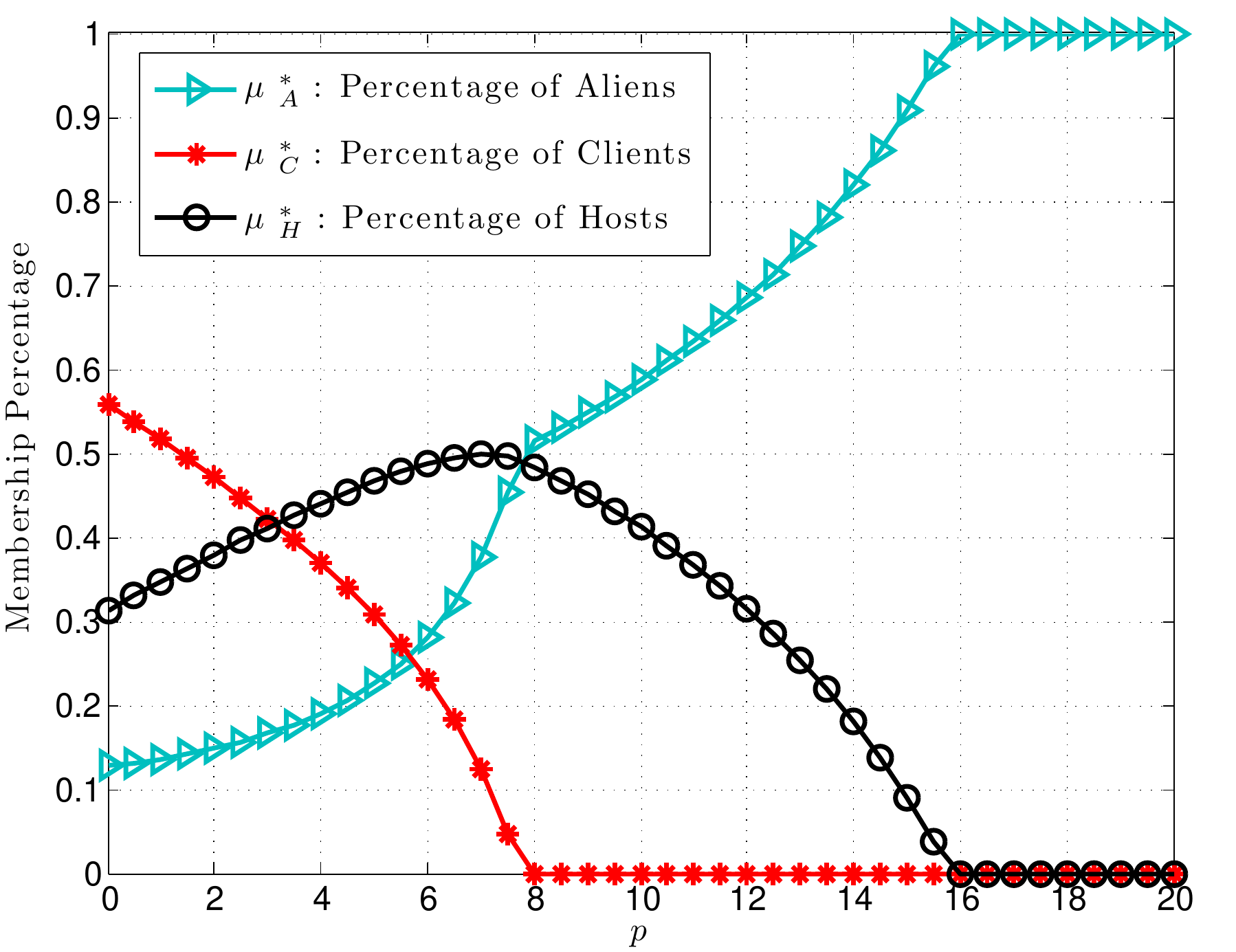}
\caption{   Equilibrium   \{$\mu^*_\a$, $\mu^*_\c$, $\mu^*_\h$\} vs  $p$ (when $\delta = 0.4$ and $\lambda=5$)}
\label{fig:price}
\end{minipage}
\begin{minipage}[t]{0.03 \linewidth}
~
\end{minipage}
\begin{minipage}[t]{0.3 \linewidth}
\centering
\includegraphics[height=1.65 in]{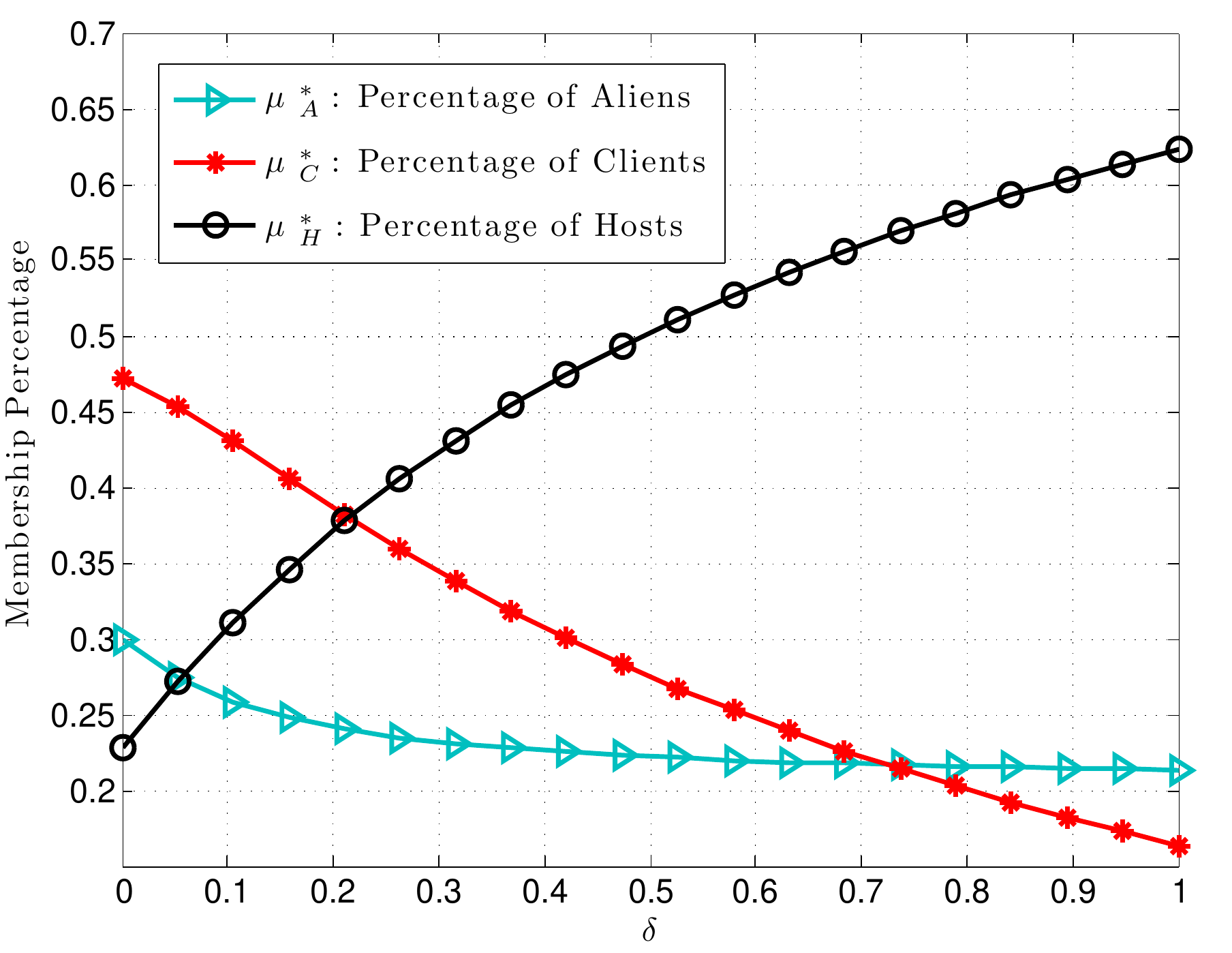}
\caption{   Equilibrium   \{$\mu^*_\a$, $\mu^*_\c$, $\mu^*_\h$\} vs    $\delta$ (when $p=5$ and $\lambda=5$)}
\label{fig:delta}
\end{minipage}
\vspace{-3mm}
\end{figure*}

\newpage

\bibliographystyle{abbrv}

\section{Appendix -- Simulations, Illustrations, and Notations Table}

\subsection{Simulations}\label{app-simulations}

We perform numerical studies in a network with the following parameters: 
$\bar{v}_{\h}= 15, \bar{v}_{\c}= 10,$ 
$\Ch= 5, \Cc= 1,$ 
$\gamma_\h= 0.5, \gamma_{\h\c}= 1, \gamma_{\c}= 0.1,$ 
and $\omega= 0.5$. 
Notice that we assume that a host's transmission cost for his own data ($\gamma_\h$) is smaller than that for a client's data ($\gamma_{\h\c}$), as the host not only needs to forward the client data to the MVNO's network  via the 3G/4G cellular connection, but also needs to communicate with the client through the Wi-Fi connection. 

\textbf{(1) Membership Selection Equilibrium in Stage II}

\emph{Impact of System Parameter $\lambda$:} 
Recall that $\lambda = N \cdot \rho$ denotes the average number of users that a user meets in each time slot, which is proportional to the user meeting probability $\rho$. 
\revhh{Figure \ref{fig:lambda} shows that under a fixed value of $(p=2,\delta=0.4)$, the equilibrium percentage of alien $\mu_\a^*$ decreases with $\lambda$, 
as a larger $\lambda$ implies a higher meeting probability among users, which encourages users to subscribe to the operator. 
Figure \ref{fig:lambda} also shows that among the subscribers, the percentage of host $\mu_\h^*$ decreases with $\lambda$ and the percentage of client $\mu_\c^*$ increases with $\lambda$. 
This implies that a larger  $\lambda$ provides more incentive for being a client than being a host.} 
 

\emph{Impacts of the MVNO's Decisions $p$ and $\delta$:} 
Figure  \ref{fig:price} shows that when increasing the price $p$, the equilibrium percentage of client $\mu_\c^*$ monotonically  decreases due to the increased data cost, while the equilibrium percentage of host  $\mu_\h^*$ first increases (when $p$ is small) and then decreases (when $p $ is large). 
\revhh{This is due to the fact that when $p$ is small, the total free data quota $\delta D$   is large (as the client demand $D = \bar{\theta}_{\c} \cdot   Y_{\c} $  is large), and hence a host can achieve  a larger monetary reward ($p\cdot \delta D$) when $p$ increases. 
When $p$ is large (e.g., $p\geq 8$), the free data quota $\delta D$ is small  (since the client demand $D$ becomes small), and hence the host can only achieve little monetary reward ($p\cdot \delta D$) when $p$ increases.} 
Figure \ref{fig:delta} shows that the equilibrium percentage of host $\mu_\h^*$ monotonically increases with the free data quota ratio $\delta$, and the equilibrium percentage of client $\mu_\c^*$ monotonically decreases with $\delta$. 
\revhh{This implies that a larger  $\delta$ provides more incentive for being a host than being a client.} 


\textbf{(2) MVNO's Strategy and Profit in Stage I} 

\begin{figure}[t]
\centering
\includegraphics[width=2.8 in]{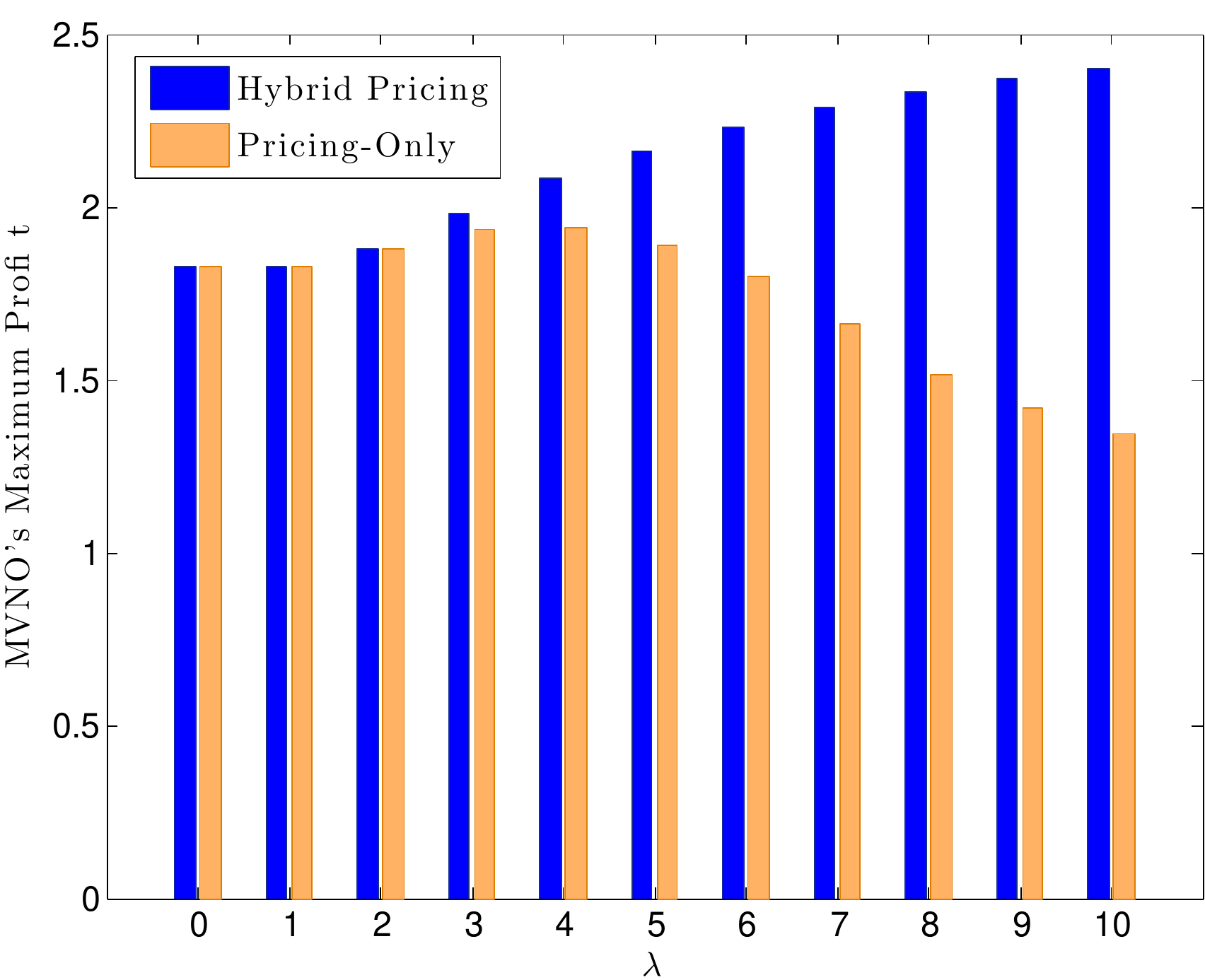}
 \caption{MVNO's maximum profit as a function of $\lambda$}
\label{fig:maxpayoff}
\end{figure}

In Figure \ref{fig:maxpayoff}, we present  the MVNO's maximum profits under the proposed best pricing and incentive mechanism (called \emph{hybrid pricing})  
and the optimal \emph{pricing-only} strategy without incentive (i.e., for a fixed $\delta = 0$). 
\revhh{Figure \ref{fig:maxpayoff} shows that as long as $\lambda>0$, our proposed hybrid pricing strategy always outperforms the pricing-only strategy in terms of the MVNO's profit.
Such a profit gain increases with $\lambda$, and can reach 50\% when $\lambda = 10$.} 
This implies that the MVNO can achieve a larger gain in a dense network with a larger $\lambda$. 

Figure \ref{fig:maxpayoff} further shows that the MVNO's profit always increases with $\lambda$ under the hybrid pricing strategy. However, the profit first increases and then decreases with $\lambda$ under the pricing only strategy. 
\revhh{This is because under our proposed hybrid pricing strategy, both hosts and clients benefit from a larger $\lambda$; while under the pricing-only strategy, only clients benefit from a larger $\lambda$ (due to the lack of incentive for hosts). 
Hence, under the hybrid pricing strategy, the MVNO can potentially generate more profit from hosts and clients with an increased value of $\lambda$. 
Under the pricing-only strategy, when $\lambda$ is small, the MVNO can generate more profit (from an increased number of clients) when $\lambda$ increases; while when $\lambda$ is large, the MVNO will generate less profit (due to a reduced number of hosts) when $\lambda$ increases. 
Intuitively, our proposed hybrid pricing strategy exploits the benefits of larger values of $\lambda$ for both hosts and clients, while the pricing-only strategy only exploits the benefit for clients.}

\begin{figure}[t]
\centering
\includegraphics[width=3 in]{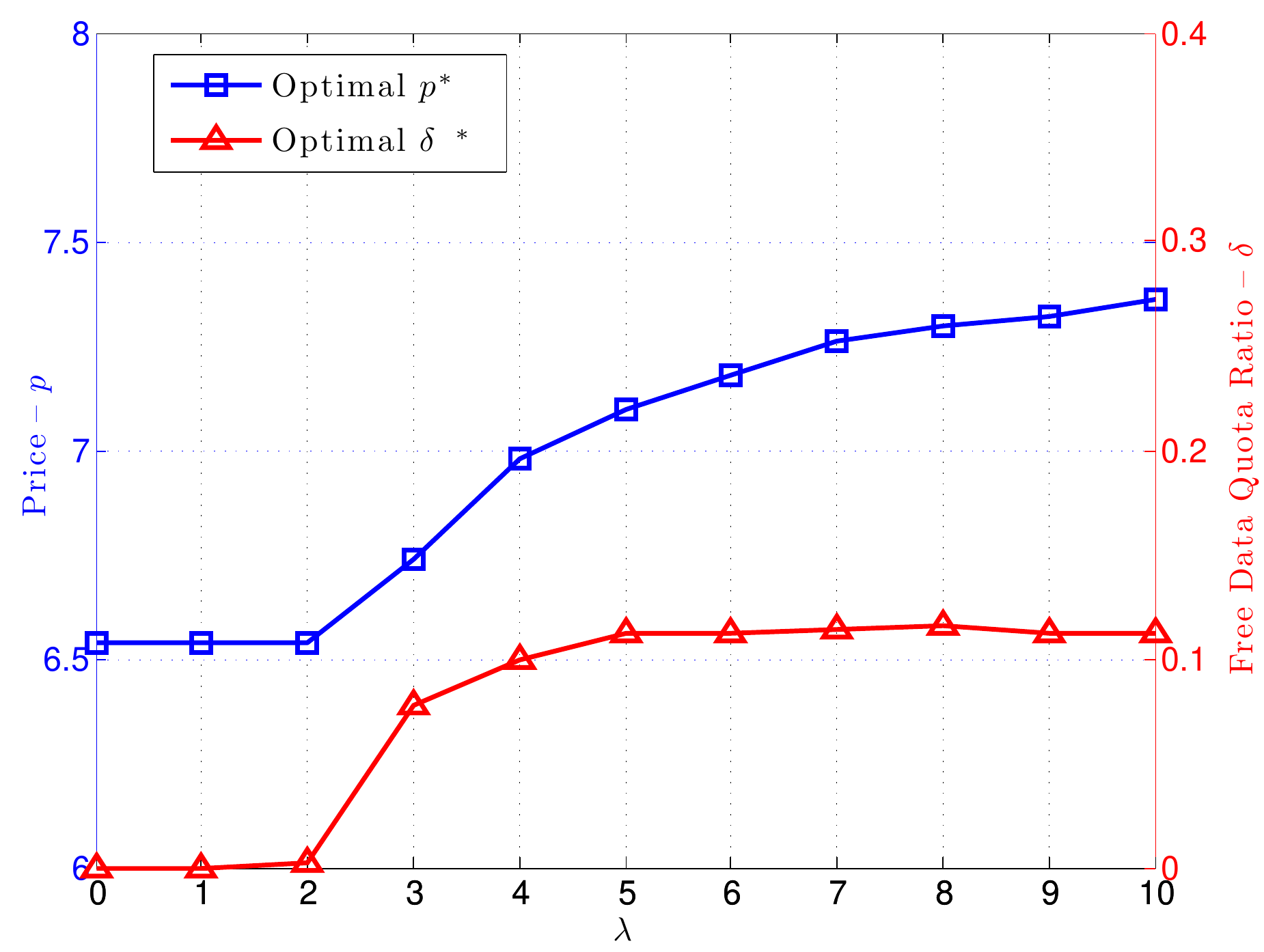}
 \caption{MVNO's best hybrid pricing strategy as a function of $\lambda$}
\label{fig:optimalPricing}
\end{figure}

Figure \ref{fig:optimalPricing} shows the MVNO's best hybrid pricing strategy under different values of $\lambda$. 
We can see that both price $p^*$ and free data quota ratio $\delta^*$ increase with $\lambda$. 
This is because both hosts and clients benefit from a larger $\lambda$, hence the MVNO can potentially charge a higher price to both hosts and clients. 
Due to the increased revenue from hosts and clients, the MVNO is willing to award more free data quota to hosts, so as to attract more hosts (and further attract more clients).

\subsection{Illustration of Equilibrium in Section \ref{sec:analysis-2}}\label{app-equi}

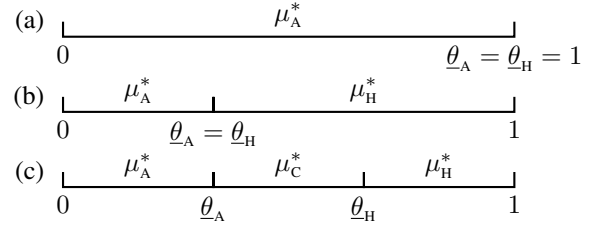
\begin{figure}[t]
\centering
\begin{tikzpicture} [scale = 1]
\draw [thick] (0,2.2) -- (0,2) -- (6,2) -- (6,2.2);
\draw  [thick] (0,1.2) -- (0,1) -- (2, 1) -- (2, 1.2) -- (2, 1) -- (6,1) -- (6,1.2);
\draw [thick]  (0,0.2) -- (0,0) -- (2, 0) -- (2, 0.2) --  (2, 0) --  (4, 0) --  (4, 0.2) --  (4, 0) --  (6,0) -- (6,0.2);
\draw (0,2.2) node[anchor = east] {(a)~~}
	  (3, 2)  node[anchor = south] {$\mu_\a^*$}
	  (0, 2)  node[anchor = north] {$0$}
	  (6, 2)  node[anchor = north] {$\underline{\theta}_{\a }  = \underline{\theta}_{\h} = 1$}
	  (0,1.2) node[anchor = east] {(b)~~}
	  (0, 1)  node[anchor = north] {$0$}
	  (6, 1)  node[anchor = north] {$1$}
	  (1, 1)  node[anchor = south] {$\mu_\a^*$}
	  (4, 1)  node[anchor = south] {$\mu_\h^*$}
	  (2, 1)  node[anchor = north] {$\underline{\theta}_{\a }  = \underline{\theta}_{\h}$}
	  (0,0.2) node[anchor = east] {(c)~~}
	  (0, 0)  node[anchor = north] {$0$}
	  (6, 0)  node[anchor = north] {$1$}
	  (1, 0)  node[anchor = south] {$\mu_\a^*$}
	  (3, 0)  node[anchor = south] {$\mu_\c^*$}
	  (5, 0)  node[anchor = south] {$\mu_\h^*$}
	  (2, 0)  node[anchor = north] {$\underline{\theta}_{\a }$}
	  (4, 0)  node[anchor = north] {$\underline{\theta}_{\h }$};
\end{tikzpicture}
\caption{Three types of equilibrium: (a) All Aliens, (b) Aliens and Hosts, (c) Aliens, Clients, and Hosts.}
\label{fig:eqtype}
\end{figure}

Figure \ref{fig:eqtype} illustrates three different types of membership selection equilibrium: 
(a) All users choose to be aliens (where $\underline{\theta}_{\a } = \underline{\theta}_{\h } = 1$, and $\mu_\a^* = 1, \mu_\c^* = \mu_\h^* = 0$). This occurs, for example, when both the time-average costs of being a client $\Cc$ and being a host $\Ch$ are very large; (b) Low type users choose to be aliens, high type users choose to be hosts, and no user chooses to be clients (where $\underline{\theta}_{\a } = \underline{\theta}_{\h } < 1$, and  $\mu_\a^* > 0, \mu_\c^* = 0, \mu_\h^* >0$). This occurs, for example, when $\Cc$ is close to $\Ch$; and (c) Low type users choose to be aliens, medium type users choose to be clients, and high type users choose to be hosts (where $\underline{\theta}_{\a } < \underline{\theta}_{\h } < 1$ and $\mu_\a^* > 0, \mu_\c^* > 0, \mu_\h^* >0$).

\subsection{Illustration of Membership Dynamics in Section \ref{sec:analysis-2}}\label{app-dynamics}

\begin{figure}[t]
\centering
\includegraphics[width=3in]{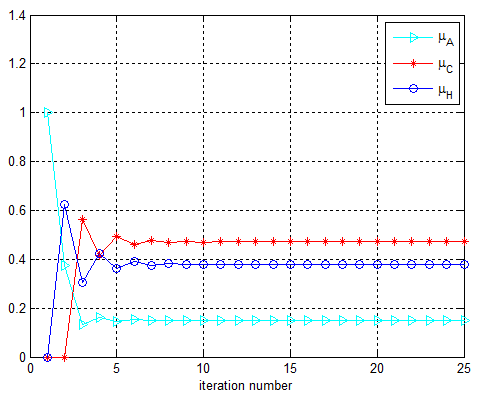}
\caption{Dynamics of $\mu_\a$, $\mu_\c$, and $\mu_\h$.}
\label{fig:dynamics}
\end{figure}

Figure \ref{fig:dynamics} illustrates the dynamics of membership percentages $\mu_\c$ and $\mu_\h$ under a given network setting. 
In this example, the membership dynamics converge to the equilibrium within around 10 iterations, and the equilibrium membership percentage is $(\mu_\c = 0.47, \mu_\h = 0.38, \mu_\a = 0.15)$. 

\subsection{Illustration of $\delta^*(p)$ and $p^*(\delta)$ in Section \ref{sec:analysis-1}} \label{app-curve}

\begin{figure}[t]
\centering
\includegraphics[width=3.2 in]{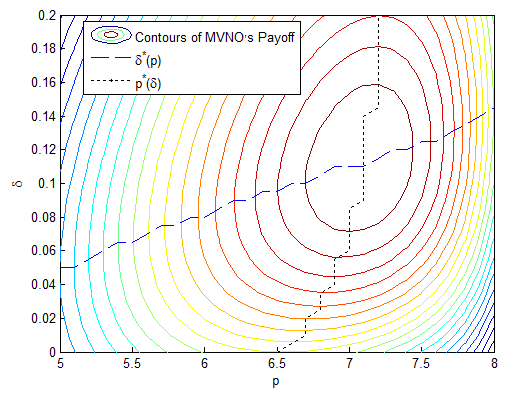}
\caption{Illustration of $\delta^*(p)$ and $p^*(\delta)$.}
\label{fig:contour}
\end{figure}

Figure \ref{fig:contour} illustrates the optimal $\delta^*(p)$ curve (blue) and the optimal $p^*(\delta)$ curve (black). 
The optimal MVNO strategy $(p^*,\delta^*)$ occurs at the intersection point of $p^*(\delta)$ and $\delta^*(p)$. 
In this example, the optimal MVNO strategy is $(p^*,\delta^*)=(7.1,0.1)$.
 The pricing only case corresponds to $p^\ast(0)$, i.e., $p=6.5$ and $\delta=0$. 

\subsection{Notations Table}\label{app-notation}

The key notations in this paper are listed in Table \ref{tbl:notation}. 

\begin{table}[t]
\caption{Key Notations}\label{tbl:notation}
\resizebox{1.22\textwidth}{!}{\begin{minipage}{\textwidth}
\begin{tabular}{| l l |}
\hline
\multicolumn{2}{|l|}{\textbf{MVNO-Related Variables}}\\
\hline
$p$          & The price (per byte of data) for hosts and clients \\
$\delta$     & The free data quota ratio for hosts \\
\hline
\multicolumn{2}{|l|}{\textbf{User-Related Parameters}}\\
\hline
$N$    	& The number of users \\
$\rho$    & Encountering (meeting) probability of  users \\
$\lambda$  & Average number of users that a user meets, $\lambda =N\rho$ \\
$\theta$  & Service request probability of users \\
$\Cc$  & Time-average cost of being client \\
$\Ch$   & Time-average cost of being host \\
$\bar{v}_{\c}$ & Average data value  for clients \\
$\bar{v}_{\h}$  & Average data value for hosts \\
$\costc$  & Average transmission cost of clients \\
$\costh$  & Average transmission cost of hosts (for own data) \\
$\costhc$ & Average transmission cost of hosts (for client data) \\
$\Pi_\c$  & Average benefit of clients for consuming one byte \\
          & of data, $\Pi_\c = \bar{v}_{\c} - \costc  - p$   \\
$\Pi_\h$  & Average benefit of hosts for consuming one byte   \\
          & of data, $\Pi_\h = \bar{v}_{\h} - \costh - p\cdot (1-\delta)$  \\ 
$\widetilde{\Pi}_\h$ & Average benefit of hosts for forwarding one byte \\
          & of data for clients, $\widetilde{\Pi}_\h = \delta\cdot p- \costhc$   \\
\hline
\multicolumn{2}{|l|}{\textbf{Important Variables}}\\
\hline
$\mu_{\a}$    & Percentage of aliens in the network \\
$\mu_{\c}$    & Percentage of clients in the network \\
$\mu_{\h}$    & Percentage of hosts in the network \\
$ \bar{\theta}_{\c}$ & Average service request probability of clients \\
$ \bar{\theta}_{\h}$ & Average service request probability of hosts \\
$ P_{\h}$  & Probability of meeting at least one host \\ 
$Y_{\c}$ & Average number of clients connected to a host \\
\lasthline
\end{tabular}
\end{minipage} }
\end{table}

\end{document}